%% file: main.tex
\documentclass{article}

\usepackage{PRIMEarxiv}

\usepackage[utf8]{inputenc}
\usepackage[T1]{fontenc}
\usepackage{hyperref}
\usepackage{url}
\usepackage{booktabs}
\usepackage{amsfonts}
\usepackage{amsmath}
\usepackage{amssymb}
\usepackage{nicefrac}
\usepackage{microtype}
\usepackage{listings}
\usepackage{fancyhdr}
\usepackage{graphicx}
\usepackage{adjustbox}
\usepackage{caption}
\graphicspath{{media/}}

\newcommand{\Emin}{E_{\min}}
\newcommand{\N}{N_0}

\title{Tensor Network based HOBO Solver}

\author{
  Yuichiro Minato \\
  blueqat Inc. \\
  Shibuya2-24-12, Shibuya \\
  Tokyo, Japan\\
  \texttt{minato@blueqat.com} \\
}

\begin{document}
\maketitle

\begin{abstract}
Higher-order binary optimization (HOBO) problems are usually reduced to quadratic (QUBO) form before they are handed to a solver. The first version of this paper proposed to keep the higher-order terms as one coefficient tensor, evaluate the cost by tensor contraction (\texttt{einsum}) and search with simulated annealing (SA), but gave no experiments. This revised version turns the same cost function into a working solver and reports what a tensor-network (TN) formulation can do that a heuristic search does not: certify, count and sample. Writing the cost as a factor graph---one small table per term or clause, one COPY tensor per variable---the contraction of that network in different semirings returns the exact minimum (tropical), the ground-state degeneracy $\N$ (counting tropical), the low-energy spectrum (truncated polynomial), the partition function (log-sum-exp) and exact Boltzmann samples (a top-down pass), all from one contraction tree found by a hyper-optimised path search with slicing. We measure the reach of exact contraction on six problem families (planted 3-regular 3-XORSAT to $n{=}150$, random 3-spin glasses, random 3-SAT, LABS, and two-dimensional plaquette and triangle models to $n{=}144$), counting up to $4.2\times10^{10}$ ground states of the $n{=}144$ lattices in minutes on a CPU (the $n{=}150$ XORSAT instance took 7.3 hours on one GPU). Two findings concern the formulation itself. First, the coefficient tensor and the term-wise factor network contract to identical exact answers, but relaxations built on the term-wise network give tighter lower bounds (by up to an order of magnitude on LABS), which turn SA upper bounds into certified optimality gaps and, on 14 of our 44 larger test instances, into certificates of optimality. Second, using the exact thermodynamics as a yardstick, we find that the loss of SA performance after quadratization cannot be attributed to the equilibrium weight of the ground states alone: at a common temperature the exact ground-state probability of the quadratized model stays within a factor of two of the native value (about three on LABS), while SA's success rate drops by up to two orders of magnitude and in several runs to zero; and, in 10 of our 12 test instances (8 in a first pass, 10 after rerunning the largest four with more effort), SA's landing distribution over degenerate ground states becomes measurably more biased after quadratization. We are explicit about what does not work: dense random instances, LABS and branch-and-bound with the bounds at hand gain nothing from the TN, SVD-compressed contraction estimates but never certifies, and we make no claim that tensor networks outperform SA as optimizers.
\end{abstract}


\section{Introduction}
\label{sec:intro}

Combinatorial optimization problems handed to quantum annealers, Ising machines and their classical emulators are usually written as Quadratic Unconstrained Binary Optimization (QUBO)~\cite{Perdomo-Ortiz2012,Glos2022,Andrew2014,Suksmono2019,Suksmono2022}. Many problems, however, are naturally of higher order---satisfiability, factoring, autocorrelation, $k$-body spin glasses---and reducing them to quadratic form costs auxiliary variables and penalty terms~\cite{boros2002pseudoboolean,rosenberg1975reduction,mandal2020compressed}. The first version of this paper~\cite{minato2024hobo} proposed to skip the reduction: keep the Higher-Order Binary Optimization (HOBO) cost as one coefficient tensor, evaluate it by tensor contraction, and search with simulated annealing (SA). A companion implementation, HOBOTAN~\cite{yasuda2024hobotan}, showed small worked examples. Neither contained a systematic experiment.

Since then the question ``is it better to keep higher-order terms than to quadratize?'' has been answered elsewhere, largely in the affirmative: with tensor-sampling optimisers~\cite{salloum2026hubo}, with matrix-product-operator SAT solvers~\cite{zec2026tnsat}, and with higher-order simulated bifurcation~\cite{kanao2023sbhigher}. The related question ``do tensor networks beat annealing as optimizers?'' has also been examined and the answer is, for now, no: on Max-3-SAT simulated annealing matched or outperformed DMRG on all instances tested~\cite{zec2026tnsat}, and on large quasi-two-dimensional random spin glasses a tensor-network branch-and-bound was 0.1--1\% worse and two orders of magnitude slower than Ising machines, though better on planted instances~\cite{dziubyna2025limitations}. Those comparisons use their own hardware and time budgets, but a revision of~\cite{minato2024hobo} that simply claimed speed would have to argue against them, and we do not.

This version keeps the proposal of v1 and adds what it lacked: an implementation, experiments, and comparisons; where the measurements do not support the tone of v1, the claims have been changed to match them. It asks a different question: given that the HOBO is written as a tensor network anyway, \emph{what can the network do that a heuristic cannot?} Three things, it turns out, each backed by experiments in Sec.~\ref{sec:experiments}.
\begin{enumerate}
\item \textbf{One network, many answers.} Contracting the factor network in the tropical $(\min,+)$ semiring gives the exact minimum~\cite{liu2021tropical}; in the counting-tropical semiring, the minimum and its degeneracy; in a truncated-polynomial semiring, the number of states at each of the lowest energy levels~\cite{liu2023generic}; in the $(\mathrm{logsumexp},+)$ semiring, the partition function; and a top-down pass through the same tree returns a minimizer or exact Boltzmann samples. The contraction path and the slicing are found once by a hyper-optimised search~\cite{gray2021hyper,cotengra} and reused for every semiring.
\item \textbf{The formulation matters for bounds, not for exact answers.} The coefficient-tensor form of v1 and the term-wise factor form contract to the same exact quantities, but a width-bounded relaxation of the term-wise network (mini-bucket elimination~\cite{dechter2003minibuckets} after a dual-ascent reparameterisation) gives tighter lower bounds than the same relaxation of the coefficient-tensor network (up to an order of magnitude on LABS, none on 3R3X). Combined with an SA upper bound this certifies optimality gaps on instances far beyond exact reach.
\item \textbf{A yardstick for heuristics.} Because the exact equilibrium quantities are available, the behaviour of SA can be measured against them. Two results follow: the collapse of SA after quadratization is not accompanied by a comparable change of the equilibrium ground-state weight, which points to the dynamics of single-bit moves as the cause, and SA does not sample degenerate ground states uniformly---and less so after quadratization.
\end{enumerate}

We keep the structure of v1: Secs.~\ref{sec:formulation}--\ref{sec:tensor} restate the HOBO coefficient tensor, Sec.~\ref{sec:contraction} goes from the single \texttt{einsum} of v1, which evaluates one assignment, to semiring contraction of a factor network over all assignments along an optimised tree, Sec.~\ref{sec:bounds} adds relaxations and moves the SVD discussion of v1 to where it belongs, Sec.~\ref{sec:sa} describes the SA baseline and quadratization, Sec.~\ref{sec:experiments} reports the experiments, and Sec.~\ref{sec:discussion} lists what did not work. The code is available (Sec.~\ref{sec:code}).

\section{Formulation of QUBO and HOBO Problems}
\label{sec:formulation}

A QUBO cost is a quadratic polynomial in binary variables $x\in\{0,1\}^n$,
\begin{equation}
Q(x) = \sum_{i} c_i x_i + \sum_{i<j} q_{ij} x_i x_j ,
\end{equation}
into which constraints are folded with penalty weights. A HOBO cost extends this to higher-order monomials,
\begin{equation}
E(x) = c_0 + \sum_{S\subseteq[n]} c_S \prod_{i\in S} x_i ,
\label{eq:hobo}
\end{equation}
where $S$ runs over index sets; since $x_i^2=x_i$, repeated indices carry no information and the polynomial is multilinear. The order $k=\max|S|$ of a problem written directly is typically small (3 for 3-SAT and 3-spin glasses, 4 for autocorrelation and plaquette models), while its quadratization introduces one auxiliary bit per eliminated pair of variables~\cite{rosenberg1975reduction,boros2002pseudoboolean}.

\section{QUBO Matrix and HOBO Tensor}
\label{sec:tensor}

\subsection{The coefficient tensor (v1)}
In v1 the coefficients of Eq.~\eqref{eq:hobo} were arranged into one order-$k$ tensor $T$ with $c_S$ stored at $T_{s_1\cdots s_k}$, the index list being $S$ padded by repeating its first element ($T_{iii}=c_i$, $T_{iij}=q_{ij}$, $T_{ijk}=h_{ijk}$ for $k=3$), so that
\begin{equation}
E(x) = c_0 + \sum_{i,j,k} T_{ijk}\, x_i x_j x_k = c_0 + \texttt{einsum}(\texttt{'ijk,i,j,k->'},\,T,\,x,\,x,\,x).
\label{eq:einsum}
\end{equation}
This is the QUBO matrix generalised by one index per order. We keep it as the \emph{definition} of the cost and as a convenient evaluation for small $n$; it is dense ($n^k$ entries) and it does not know which monomials came from the same term of the original problem.

\subsection{The factor network}
\begin{figure}[t]
\centering
\includegraphics[width=0.5\linewidth]{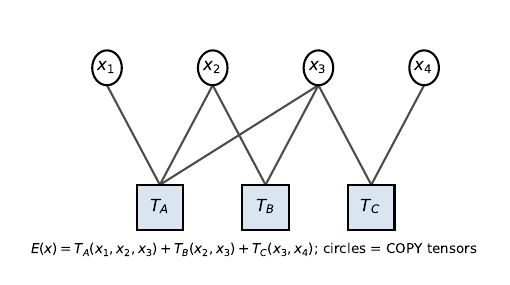}
\caption{The factor network of a HOBO with three terms on four variables. Squares are energy tables, circles are COPY tensors (hyperedges). Contracting the network in a semiring sums the tables over all assignments in that semiring's sense.}
\label{fig:tn}
\end{figure}

For contraction we use a different network, built from the sparse form that the problem itself provides (Fig.~\ref{fig:tn}). Every term---a clause of a SAT formula, a $k$-spin coupling, an autocorrelation product---becomes a small table $T_a$ over $\{0,1\}^{|S_a|}$ holding that term's energy for each assignment of its variables, and every variable becomes a hyperedge shared by the tables that mention it (a COPY or $\delta$ tensor). The cost is the sum of the tables,
\begin{equation}
E(x) = c_0 + \sum_a T_a(x_{S_a}),
\label{eq:factor}
\end{equation}
which is a tensor network in the $(+,\times)$ sense once each table is exponentiated, and directly a tensor network in the $(\min,+)$ or $(\mathrm{logsumexp},+)$ sense (Sec.~\ref{sec:contraction}).

Three ways of cutting the same function $E$ into tables will be compared. \emph{Bit monomials}: one table per maximal index set of Eq.~\eqref{eq:hobo}, lower-order monomials absorbed into a maximal set containing them---this is what the coefficient tensor of v1 induces. \emph{Spin monomials}: the same after the change of variables $s_i=1-2x_i$. \emph{Natural factors}: the tables of the generating problem (one per clause, one per coupling, with the polynomial expansion never carried out). The three contract to identical exact answers; they differ as starting points for relaxations (Sec.~\ref{sec:exp-basis}).

\section{Solving by Contraction: One Tree, Many Semirings}
\label{sec:contraction}

\subsection{Semirings}
Let $\oplus$ and $\otimes$ be the addition and multiplication of a commutative semiring. Contracting the network of Eq.~\eqref{eq:factor} means computing $\bigoplus_{x}\bigotimes_a T_a(x_{S_a})$ with $\otimes$ acting on energies as $+$ (energies add) and $\oplus$ chosen as follows.
\begin{itemize}
\item $\oplus=\min$ (tropical): the result is $\Emin=\min_x E(x)$~\cite{liu2021tropical}.
\item $\oplus$ on pairs $(E,C)$: $(E_1,C_1)\oplus(E_2,C_2)$ keeps the smaller energy and sums the counts of equal energies; $\otimes$ adds energies and multiplies counts. The result is $(\Emin,\N)$ with $\N$ the ground-state degeneracy~\cite{liu2021tropical,liu2023generic}.
\item $\oplus$ on $(E,\mathbf{c})$ with $\mathbf{c}\in\mathbb{R}^K$: all energies are integer multiples of a unit $u$ ($u=1$ for SAT, $2$ for $\pm1$ spin models, $4$ for LABS), and $\mathbf{c}_j$ is the number of states at energy $E+ju$, $j<K$. $\oplus$ aligns two elements to the smaller $E$ and adds the shifted count vectors; $\otimes$ adds energies and convolves the count vectors; entries beyond $E+(K-1)u$ are discarded. The result is the exact number of states at each of the $K$ lowest levels (the truncated polynomial of~\cite{liu2023generic}).
\item $\oplus=\mathrm{logsumexp}$ on $-\beta E$: the result is $\log Z(\beta)$ (every reduction subtracts its maximum before exponentiating, so large $\beta$ causes no overflow).
\end{itemize}
A top-down pass through the same binary tree recovers a minimizer or draws exact samples. For each pairwise contraction we keep the combined tensor $S$ before its summed indices are reduced; walking the tree from the root, the indices reduced at a node are drawn (or chosen as the argmin) from $S$ restricted to the values already assigned to that node's output indices---for sampling, with probability proportional to $e^{S}$, where $S$ is the sum of the two children's log-partial-partition tensors (the contraction being run on $-\beta E$); this is the exact conditional of those indices given the ones above. Every quantity above is exact for what it defines: the contraction introduces no approximation, and the only truncation is the deliberate cut at $K$ levels of the truncated polynomial.

\subsection{Contraction order, slicing and execution}
The cost of contracting a network is exponential in the width of the contraction tree (the largest intermediate tensor), so the tree matters more than the hardware. We use \texttt{cotengra}~\cite{gray2021hyper,cotengra}: a hyper-optimiser over greedy and hypergraph-partitioning (KaHyPar) path finders with subtree reconfiguration, minimising the total contraction cost, followed by slicing (fixing a few indices and summing over their values) so that the largest intermediate fits in memory. Slicing preserves the result and splits the work into independent pieces, which parallelise trivially; it can, however, increase the total cost, sometimes severely (Sec.~\ref{sec:discussion}). This is the same kind of tree that NVIDIA's \texttt{cuTensorNet}~\cite{cutensornet} plans; since, to our knowledge, \texttt{cuTensorNet} and \texttt{cuTENSOR} execute contractions in $(+,\times)$ only, we plan the path with \texttt{cotengra} and execute each pairwise contraction ourselves as a batched min-plus (or counting) matrix product, chunked over the summed index so that memory stays at the size of the output. This runs on CPU (NumPy or PyTorch) and on GPU (PyTorch, Sec.~\ref{sec:exp-xorsat}). The tree is chosen by the size of the intermediate tensors, which does not depend on the semiring, so a tree found for one semiring can be reused for the others; the constant factors differ (the counting semiring carries a second array, the truncated polynomial $K$ entries per element, the sampler keeps the unreduced intermediates), but the exponential part of the cost is the same. We stress what is and is not carried over from v1. The network of v1 is the coefficient tensor joined to copies of the vector $x$: its indices run over the variable labels $1,\dots,n$ and its contraction evaluates the cost of one given assignment. The factor network contracted here is a different network built from the same coefficients: its indices are the variables themselves, each of dimension two, and its contraction runs over all $2^n$ assignments. What is kept from v1 is the principle---the cost is a tensor network and quantities of interest are obtained by contracting it---not the network.

For comparison we also report plain variable elimination in the order given by the min-fill heuristic, which is what a bucket-elimination implementation would do without a path optimiser.

\section{Relaxations: Certified Bounds and Approximate Contraction}
\label{sec:bounds}

\subsection{Lower bounds by cutting hyperedges}
Exact contraction is out of reach when the width exceeds roughly 33 on a CPU (Sec.~\ref{sec:exp-reach}). Mini-bucket elimination~\cite{dechter2003minibuckets} bounds the width: when eliminating a variable $v$ would create a table over more than $i+1$ variables, the tables that share $v$ are split into groups and $v$ is minimised in each group separately. In tensor-network language, the hyperedge $v$ is cut and each group receives its own copy of $v$; since $\min_v(f+g)\ge\min_v f+\min_v g$, the contraction value is a valid lower bound on $\Emin$, exact once $i$ reaches the width. Before the split we equalise the min-marginals of the groups (moment matching, in the spirit of~\cite{liu2011holder,ahn2019bucket}); this keeps the bound valid and usually tightens it, but not always---downstream combination can undo the gain, and we found small counterexamples---so we compute the bound with and without it. We first run a dual-ascent reparameterisation on the whole network (min-sum diffusion: shift energy between a variable's tables so that their min-marginals over that variable agree; the sum of the tables is unchanged and $\sum_a\min T_a$ is the dual of the local-polytope LP relaxation), then mini-bucket on the reparameterised tables. Every number produced---the dual-ascent value at any iteration, and the mini-bucket value at any $i$-bound with or without moment matching---is a valid lower bound, and the reported LB is the maximum of them.

Together with any feasible configuration---we use the best of an SA run---this yields a certified optimality gap $\mathrm{UB}-u\lceil\mathrm{LB}/u\rceil$, where $u$ is the energy unit of the family (within an instance all energies differ by integer multiples of $u$, and in every instance used here the energies themselves are multiples of $u$, so rounding the bound up to the next multiple keeps it valid; $u=2$ for $\pm1$ spin models and $u=1$ for 3-SAT) on instances where exact contraction is impossible.

\subsection{Approximate contraction (the SVD of v1)}
v1 discussed lightening heavy tensors by singular value decomposition. The precise version of that idea is compressed contraction~\cite{pan2020contracting,gray2024hyperoptimized}: contract in $(+,\times)$ at finite $\beta$ and truncate intermediate bonds to a maximal dimension $\chi$ by SVD. We tested it with \texttt{quimb}'s \texttt{contract\_compressed} on the two-dimensional plaquette model (Sec.~\ref{sec:exp-svd}). It estimates $\log Z$ well at moderate $\chi$, but the estimate is neither an upper nor a lower bound, and SVD truncation has no meaning in the tropical semiring. It is therefore a tool for free-energy estimates, not for certificates, and we do not use it further.

\section{Simulated Annealing and Quadratization}
\label{sec:sa}

The SA of v1 is retained as the heuristic baseline and made batched: $B$ replicas are annealed at once (PyTorch, CPU or GPU) with single-bit Metropolis moves. The flip cost of bit $i$ needs only the terms containing $i$, $\Delta E_i=(1-2x_i)\sum_{S\ni i}c_S\prod_{j\in S\setminus i}x_j$, so the variables are greedily coloured on the interaction graph and all bits of one colour (which share no term) are updated in parallel; such bits are conditionally independent given the rest, so the joint update is a product of single-bit Metropolis moves and preserves detailed balance. The schedule is geometric in $\beta$ from $\beta_{\rm hot}=\ln 2/\max_i A_i$ to $\beta_{\rm cold}=\ln 100/\min_S|c_S|$, $A_i=\sum_{S\ni i}|c_S|$, a rule modelled on the default temperature range of D-Wave's \texttt{neal}~\cite{dwaveneal} (hot enough to flip the most strongly coupled bit with probability one half, cold enough to suppress the smallest excitation a hundredfold). A quadratized instance is just another HOBO of order two with more bits, so native and quadratized runs use the same kernel and the same schedule rule.

Quadratization is Rosenberg's substitution~\cite{rosenberg1975reduction}: repeatedly pick the pair $(a,b)$ shared by the most terms of order $\ge 3$, introduce $y=x_ax_b$, replace the pair by $y$ in those terms and add the penalty $M(x_ax_b-2x_ay-2x_by+3y)$, which is zero iff $y=x_ax_b$ and at least $M$ otherwise. With $A$ the sum of $|c_S|$ over the terms that received $y$, any $M>A$ is exact; for integer coefficients $M=A+1$ is the tightest exact integer choice and we also test $M=1.5A$.

\section{Experiments}
\label{sec:experiments}

\subsection{Problem families}
All generators are in the code repository and produce integer energies (coefficients are integers, so the tie tolerance of $10^{-9}$ in the counting semiring is never exercised). The exact sampler is validated against enumeration (below) and used in Sec.~\ref{sec:exp-fair}; we report no other sampling experiment.
\emph{Random 3-spin glass}: $n$ spins, $m=2n$ random triples with $J=\pm1$, $E=\sum J_{ijk}s_is_js_k$.
\emph{Planted 3-regular 3-XORSAT (3R3X)}~\cite{kowalsky2022xorsat}: every variable in exactly three of $n$ parity clauses, right-hand sides from a planted assignment, so $\Emin=-n$ and $\N=2^{\,n-\mathrm{rank}_{\mathrm{GF}(2)}A}$ are known independently of the solver. This family is solvable in polynomial time by Gaussian elimination over GF(2); it is a standard benchmark precisely because physics-inspired heuristics fail on it.
\emph{Random 3-SAT} at clause density $4.26$, $E$ the number of violated clauses.
\emph{LABS} (low-autocorrelation binary sequences), $E=\sum_k C_k^2$, a 4-body model with known optima~\cite{packebusch2016labs}.
\emph{2D plaquette} ($4$-spin terms on the faces of an $L\times L$ torus, random signs) and \emph{2D triangle} ($3$-spin terms on the up-triangles of a periodic triangular lattice).
Before any experiment, every routine was checked against enumeration on instances with $n\le 22$ of all six families: minimum, degeneracy, level counts, $\log Z$, the recovered minimizer and the mean energy of the exact sampler agree with brute force, with and without slicing; an independent audit (a fresh enumeration script written without reference to our code) confirmed minimum, degeneracy, level counts, $\log Z$, minimizer and all one-bit sampler marginals within four standard errors on three further instances. Beyond that size the only independent checks are the planted energy and the GF(2) degeneracy of 3R3X (Sec.~\ref{sec:exp-xorsat}); for the other families the larger instances rest on the correctness of the contraction, which is exact by construction.

\subsection{Formulation: exact answers agree, bounds do not}
\label{sec:exp-basis}
Table~\ref{tab:basis} contracts the same five functions from the three table decompositions of Sec.~\ref{sec:tensor}. The exact quantities coincide: $\Emin$ and $\N$ agree exactly and $\log Z(1)$ to $4\times10^{-15}$ across the three decompositions on all five instances, and all agree with enumeration. The lower bounds do not: on LABS the trivial bound (sum of table minima) is $-5458$ from the bit-monomial tables and $-374$ from the natural factors, and at $i=8$ still $-299.7$ versus $-38$ against $\Emin=+10$; on 3-SAT the natural (clause) tables give the bound $0$ (one unit below $\Emin=1$) before any elimination, where the bit-monomial tables start at $-49$. The reason is that expanding a clause or a spin coupling into $\{0,1\}$ monomials scatters one term into many monomials of alternating sign, which land in different tables, and the cancellations then have to be recovered across tables by the relaxation. \emph{The coefficient tensor of v1 remains the right object to define the cost; the network to contract should be built from the problem's own terms.}

\begin{table}[t]
\centering\small
\caption{Lower bounds (dual ascent + mini-bucket at $i$-bound $4/8/12$, and the trivial sum of table minima) from three decompositions of the same cost. Exact contraction gives the same $\Emin$ for all three.}
\label{tab:basis}
\adjustbox{max width=\linewidth}{\input{tables/tab_basis.tex}}
\end{table}

\subsection{Reach of exact contraction}
\label{sec:exp-reach}
\begin{figure}[t]
\centering
\includegraphics[width=\linewidth]{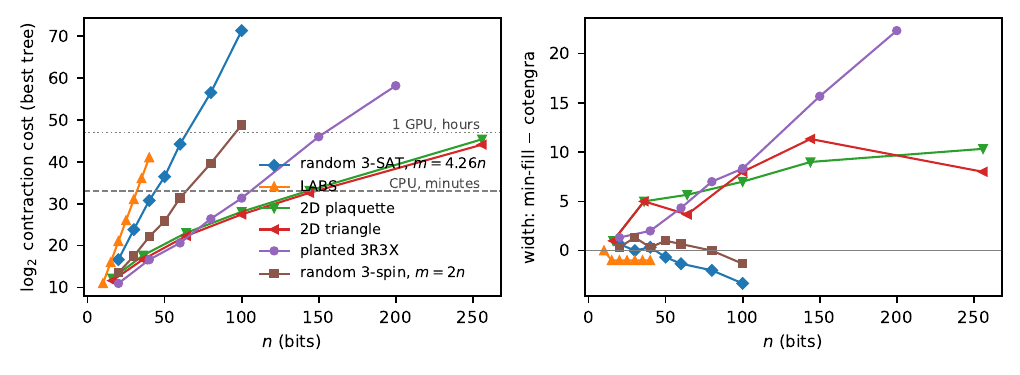}
\caption{Left: best $\log_2$ contraction cost against $n$ for the six families; the dashed lines mark what a CPU contracts in minutes and a single GPU in hours. Right: reduction of the contraction width by the hyper-optimised path search relative to min-fill elimination.}
\label{fig:reach}
\end{figure}

Table~\ref{tab:reach} and Fig.~\ref{fig:reach} give, for every family and size, the width of the min-fill elimination order and of the \texttt{cotengra} tree (30\,s of hyper-optimisation, three seeds), the best $\log_2$ contraction cost, and---where that cost is below $2^{30}$---the time of the exact counting contraction on a CPU. The path optimiser lowers the width on structured instances, by up to 26 at the largest sizes (3R3X $n{=}200$: 72--79 to 52--53; triangle $L{=}10$: 29 to 21; only 1--3 at $n\le40$), while on random 3-SAT it does not help and the 30\,s search is sometimes worse than min-fill; the cost column therefore reports the best of min-fill and the two \texttt{cotengra} searches. LABS has width $N-1$ under min-fill and $N$ in the \texttt{cotengra} trees (all pairs interact through the 4-body terms), so it gains nothing from a path search; its optima are known to $N=66$ by dedicated branch-and-bound~\cite{packebusch2016labs}, and we include it only as a negative example.

With a CPU and a few minutes, exact contraction reaches $n=100$ for 3R3X (26--190\,s, Table~\ref{tab:xorsat}), $L=12$ ($n=144$) for the two lattice models (Table~\ref{tab:lattice}), $n=50$ for random 3-spin at $m=2n$ (1.6--7.9\,s) and $n=30$ for random 3-SAT (below 1\,s; $n=40$ has cost $2^{30}$--$2^{31}$ and was not run). Table~\ref{tab:lattice} shows the $L=12$ lattices in detail: two to four minutes per instance on eight CPU threads (plus 180\,s of path search) for the exact minimum \emph{and} its degeneracy, which ranges from $6.9\times10^3$ to $4.2\times10^{10}$. On the same instances SA (1024 replicas, 2000 sweeps) finds the same minimum in every case but reaches it with $0.4$--$3.2\%$ of the replicas and, of course, cannot certify it or count it.

\begin{table}[t]
\centering\small
\caption{Width of the min-fill order and of the \texttt{cotengra} tree, best $\log_2$ contraction cost (ranges over three seeds; a single value means all seeds agree; LABS has one instance per $N$), and, where run, CPU time of the exact counting contraction and the degeneracy $\N$. ``terms'' counts the monomials of Eq.~\eqref{eq:hobo} in the $\{0,1\}$ basis, e.g.\ about $5n$ for the $n$ parity clauses of 3R3X, each of which expands into seven monomials before merging.}
\label{tab:reach}
\adjustbox{max width=\linewidth}{\input{tables/tab_reach.tex}}
\end{table}

\begin{table}[t]
\centering\small
\caption{Exact minimum and degeneracy of $L=12$ lattice models (natural factors, \texttt{cotengra} tree with 180\,s search, min-plus execution in PyTorch on 8 CPU threads; TN time excludes the search) against SA with 1024 replicas and 2000 sweeps.}
\label{tab:lattice}
\adjustbox{max width=\linewidth}{\input{tables/tab_lattice.tex}}
\end{table}

\subsection{Planted 3-XORSAT: where SA stalls}
\label{sec:exp-xorsat}
Table~\ref{tab:xorsat} compares exact contraction with SA on planted 3R3X for $n=60$ to $150$. The TN returns $\Emin=-n$ and a degeneracy that equals $2^{n-\mathrm{rank}A}$ in every instance---an independent algebraic check of the counting semiring. SA with $10^4$ sweeps reaches the planted energy with $1.8$--$3\%$ of the replicas at $n=60$, $0.3$--$0.5\%$ at $n=80$, $0$--$0.3\%$ at $n=100$, and in one $n=100$ instance not at all (best $-98$). The $n=150$ instance was contracted on one V100 GPU: 32{,}768 slices, $7.3$\,h (excluding 300\,s of path search), peak memory $5.5$\,GB, giving $\Emin=-150$ and $\N=4$, both correct, while SA on the same GPU stops at $-146$. We emphasise again that Gaussian elimination solves 3R3X in polynomial time; the point of the table is that the generic contraction, which uses no algebraic structure, reproduces the algebraic answer including the degeneracy, on a family where the heuristic of v1 does not.

\begin{table}[t]
\centering\small
\caption{Planted 3-regular 3-XORSAT. The TN degeneracy is checked against $2^{n-\mathrm{rank}_{\mathrm{GF}(2)}A}$. SA: 1024 replicas; $P(E_0)$ is the fraction of replicas that reach the planted energy. The $n=150$ instance ran on one V100 (\texttt{cotengra} search 300\,s, float32 energies, float64 counts). $^\dagger$width after slicing (the largest intermediate of one slice), which is why the $n=150$ value (26) is below the unsliced widths of Table~\ref{tab:reach}.}
\label{tab:xorsat}
\adjustbox{max width=\linewidth}{\input{tables/tab_xorsat.tex}}
\end{table}

\subsection{Anatomy of the quadratization penalty}
\label{sec:exp-quad}
\begin{figure}[t]
\centering
\includegraphics[width=\linewidth]{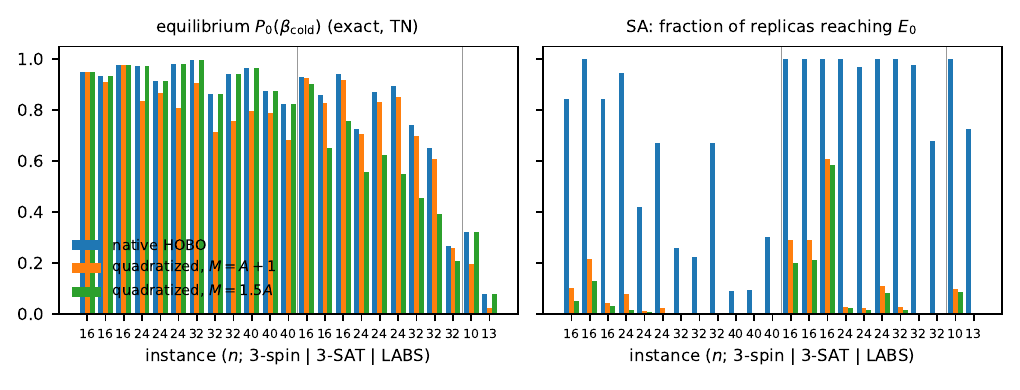}
\caption{Quadratization leaves the exact equilibrium ground-state probability nearly unchanged (left) while SA's success rate collapses (right). Instances are grouped by family (3-spin, 3-SAT, LABS) and labelled by $n$.}
\label{fig:quad}
\end{figure}

Quadratization is known to degrade solvers~\cite{zec2026tnsat,salloum2026hubo,kanao2023sbhigher}, and our SA is no exception. The exact thermodynamics lets us ask \emph{why}. For each instance we quadratize with $M=A+1$ and $M=1.5A$ and compute, at one common inverse temperature $\beta_{\rm cold}$ (that of the native schedule), the exact equilibrium probability of being in a ground state, $P_0(\beta)=\N e^{-\beta\Emin}/Z(\beta)$, for the native and both quadratized models (the auxiliary bits are summed over), and we run SA with 1000 sweeps on all three (Table~\ref{tab:quad}, Fig.~\ref{fig:quad}).

The equilibrium probabilities move by a bounded amount: on 3-spin, $P_0$ for $M=1.5A$ equals the native value to within $3\times10^{-4}$ and for $M=A+1$ is lower by at most a factor $1.25$; on 3-SAT the factors are at most $1.1$ ($M=A+1$) and $1.7$ ($M=1.5A$); on LABS $N=13$ they are $3.2$ ($M=A+1$) and $1.01$ ($M=1.5A$). At twice the inverse temperature all three models have $P_0\ge0.95$ on every 3-spin and 3-SAT instance. The degeneracy $\N$ is unchanged by construction. Yet SA's success rate falls from $0.09$--$1.0$ to at most $0.22$ and in six of the 24 quadratized 3-spin runs to zero, and from $0.68$--$1.0$ to at most $0.61$ and typically below $0.03$ for $n\ge24$ on 3-SAT, with the ratio growing with $n$, and from $0.73$ to $0$ on LABS $N=13$ where $P_0$ for $M=1.5A$ is within $1\%$ of the native value. The weight that equilibrium assigns to the ground states changes by a small factor; what changes, we conjecture, is the reachability under single-bit moves, because the auxiliary bit and its two parents must flip in a coordinated way to move between valid configurations. The comparison is suggestive rather than conclusive, and it controls some things but not everything: $P_0$ is evaluated at the same $\beta$ for all three models, each bit receives the same number of update attempts (one per sweep; the quadratized run therefore does more total work because it has more bits), but the geometric SA schedule is derived from each model's own coefficients, the auxiliary bits are annealed at the same temperature as the others, and $P_0$ summarises the spectrum at one temperature. The level counts of the quadratized models include configurations whose auxiliary bits are inconsistent with their parents; with $M=A+1$ on 3-SAT such a violation can raise the energy by as little as one unit net, which is where the extra low-lying states of Table~\ref{tab:quad} come from. These are the states a sampler of the quadratized model actually visits, so they belong in the comparison. The level counts of Table~\ref{tab:quad} (the number of states with $E\le E_0+2$, from the truncated-polynomial contraction; that is one level above the minimum for the spin models, two for 3-SAT, the ground level only for LABS) also change by less than a factor of six under quadratization, while the SA rate drops by factors from 1.6 to about 340, and to zero in nine of the 46 quadratized runs. A sampler-only method cannot make this comparison at all; the TN provides the equilibrium reference at a cost that does not depend on the sampling effort. The quadratized networks are also wider (Table~\ref{tab:quad}), so quadratization hurts exact contraction as well. This separation between equilibrium weight and dynamics is not available to a method that only samples. We note that the SA column is a ``best-so-far'' rate---the fraction of replicas whose lowest-energy configuration during the anneal, restricted to the original bits for the quadratized models, is a ground state of the original cost---and $P_0$ an instantaneous equilibrium probability; the two columns are compared only across variants, never with each other.

\begin{table}[t]
\centering\small
\caption{Native versus quadratized HOBO: number of auxiliary bits, contraction width, degeneracy, number of states with $E\le E_0+2$, exact equilibrium ground-state probability at the common $\beta_{\rm cold}$ of the native schedule and at twice that value, and SA success rate (1024 replicas, 1000 sweeps).}
\label{tab:quad}
\adjustbox{max width=\linewidth}{\input{tables/tab_quad.tex}}
\end{table}

\subsection{Does SA sample degenerate ground states fairly?}
\label{sec:exp-fair}
\begin{figure}[t]
\centering
\includegraphics[width=\linewidth]{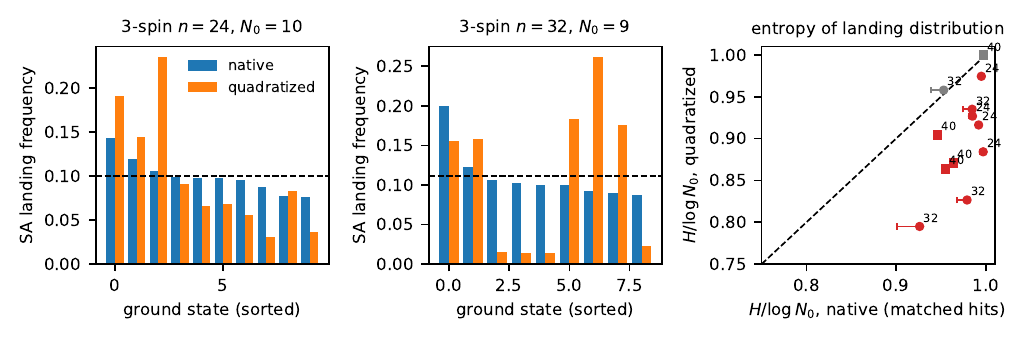}
\caption{Left, centre: SA landing frequencies over the exact ground states of two instances (dashed line: uniform). Right: normalised entropy of the quadratized landing distribution against that of the native distribution subsampled to the same number of hits (bars: 5\% quantile; squares: the $n=40$ reruns with thirteen times the quadratized effort); red points lie significantly below the diagonal.}
\label{fig:fair}
\end{figure}

When $\N>1$, a user of SA typically assumes that the ground states are found with comparable probability. This can only be checked if all ground states are known. We select random 3-spin instances with $4\le\N\le21$, count $\N$ by contraction, enumerate the ground states by exact Boltzmann sampling at large $\beta$ until $\N$ distinct ones have appeared (completeness is guaranteed by the count), and record which ground state each successful SA replica lands in ($8\times2048$ replicas; native 1000 sweeps, quadratized $M=A+1$ 3000 sweeps). Table~\ref{tab:fair} and Fig.~\ref{fig:fair} summarise twelve instances; the four $n=40$ instances were then rerun with about thirteen times the quadratized effort ($32\times4096$ replicas, 5000 sweeps; rows marked $^\dagger$). In these instances native SA reaches every ground state but not uniformly: the ratio of the most to the least frequent landing is $1.2$--$5.3$ and a $\chi^2$ test rejects uniformity in all 12 instances (first pass: $p<10^{-3}$ in 11 instances and $p=0.026$ in one, Table~\ref{tab:fair}; native hits are $2450$--$16070$ per instance, so every expected cell count exceeds 100 and the $\chi^2$ approximation is safe; replicas are independent chains driven by one generator per batch). Quadratized SA reaches ground states 5 to 120 times less often, and in the first-pass $n=40$ runs its cell counts are too small for a $\chi^2$ test; to compare the \emph{shape} of the two distributions at equal sample size we subsample the native landings, without replacement, to the quadratized hit count (1000 draws) and ask how often the subsample is at least as concentrated, measured by the normalised entropy $H/\log\N$, as the quadratized distribution. In the first pass quadratization makes the landing distribution significantly more biased (5\% level) in 8 of 12 instances: all four at $n=24$, three of four at $n=32$, and one of four at $n=40$, where the quadratized runs had only 44--82 hits. With thirteen times the effort the quadratized $n=40$ runs reach 653--1248 hits and three of the four become significant (max/min ratios 8--20 against 1.2--3.5 for the native reruns); the remaining one has $\N=4$ and both samplers are close to uniform on it. Taking the rerun values for $n=40$, 10 of the 12 instances are significant, and the same 10 survive a Benjamini--Hochberg correction at 5\% over the 12 tests; the two exceptions are the instances with $\N=21$ (not rerun) and $\N=4$. In these instances quadratization therefore costs not only success probability but also solution diversity; for the ten significant instances the entropy reduction in Table~\ref{tab:fair} is $0.02$--$0.16$ in $H/\log\N$.

\begin{table}[t]
\centering\small
\caption{SA landing distribution over the $\N$ exact ground states of random 3-spin instances ($^\dagger$: $n=40$ rerun with $32\times4096$ quadratized replicas and 5000 sweeps, about 13 times the first-pass effort). ``max/min'' is the ratio of the most to the least frequent ground state; $H/\log\N$ the normalised entropy; $\chi^2$ $p$ the $p$-value of the uniformity test on the native landings; the last two columns give the native entropy subsampled to the quadratized hit count and the fraction of subsamples at least as biased as the quadratized run ($p<0.05$: quadratized is significantly more biased).}
\label{tab:fair}
\adjustbox{max width=\linewidth}{\input{tables/tab_fair.tex}}
\end{table}

\subsection{Certified gaps beyond exact reach}
\label{sec:exp-gap}
\begin{figure}[t]
\centering
\includegraphics[width=\linewidth]{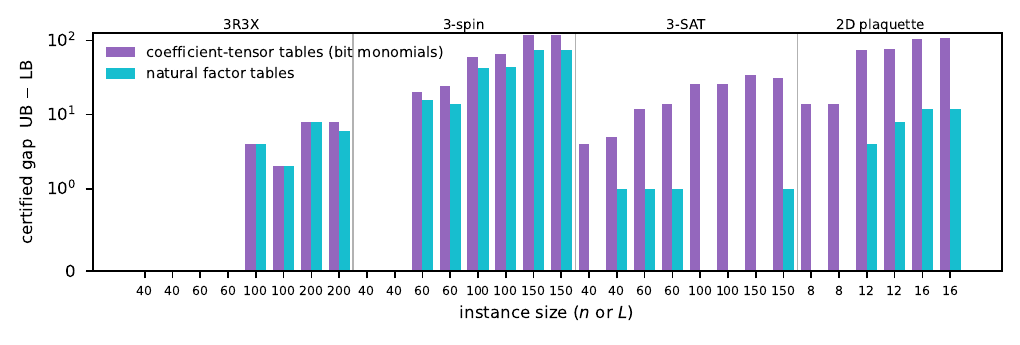}
\caption{Certified gap per instance from the two table decompositions (zero height = certified optimal; symlog scale).}
\label{fig:gap}
\end{figure}

Table~\ref{tab:gap} and Fig.~\ref{fig:gap} pair an SA upper bound (1024 replicas, 2000 sweeps) with the lower bound of Sec.~\ref{sec:bounds} at $i$-bounds up to 20 on instances up to $n=576$. With natural factors, 14 of 44 instances are certified optimal without a full contraction (in five of them the $i=20$ mini-bucket happened to need no split and was therefore exact): all 3R3X with $n\le60$ (there the sum of clause minima, $-n$, is already the planted optimum, so the certificate only requires SA to reach it), satisfiable 3-SAT instances up to $n=150$ (the clause tables give $\mathrm{LB}=0$ directly), 3-spin at $n=40$, and both lattices at $L=8$. With the coefficient-tensor tables only 6 of the same 30 instances are certified (12 with natural factors), and the gap is never smaller and larger in 21 of the 30. For the $L=12$ lattices the remaining gaps of 4--8 were resolved by the exact contraction of Table~\ref{tab:lattice}: the SA upper bound was optimal and the bound was loose. For dense random instances (3-spin $n\ge60$, 3R3X $n\ge100$) the dual ascent does not move from the trivial bound (``dual'' column of Table~\ref{tab:gap}; 3-spin $n=100$: trivial and dual $-200$, mini-bucket at $i=20$ $-192$, SA $-150$) and mini-bucket closes only part of the gap; on 3-spin the relaxation is then not informative, while on 3R3X $n\ge100$ it leaves gaps of 2--8 that SA, not the bound, is responsible for (the bound $-n$ is tight).

\begin{table}[t]
\centering\small
\caption{Certified optimality gaps $\mathrm{UB}-u\lceil\mathrm{LB}/u\rceil$ ($u=2$ for the spin and parity models, $u=1$ for 3-SAT) from an SA upper bound and lower bounds with natural-factor tables (``dual'': dual ascent alone; ``LB'': best of dual ascent and mini-bucket at $i\le20$) and with coefficient-tensor tables. $^*$: certified optimal.}
\label{tab:gap}
\adjustbox{max width=\linewidth}{\input{tables/tab_gap.tex}}
\end{table}

\subsection{Compressed contraction}
\label{sec:exp-svd}
On the plaquette model at $L=6$, \texttt{contract\_compressed} with $\chi=64$ reproduces $\log Z$ to relative error $10^{-6}$ or better at $\beta=0.3,1,3$; $\chi=16$ to $3\times10^{-4}$; $\chi=4$ overestimates $\log Z$ by $5.5\%$ at $\beta=3$. The error has either sign ($\chi=4$ underestimates at $\beta=0.3$ and $1$), so the estimate is accurate but bounds nothing, and each contraction took 8--22\,s against 5\,ms for the exact contraction at this size. At $L=8$ our first attempt ran out of memory because the hyperedge resolution produced an unfavourable path; we did not pursue it.

\section{Discussion and Limitations}
\label{sec:discussion}

\paragraph{Where the tensor network adds nothing.} Exact contraction is exponential in the contraction width. For random 3-SAT and random 3-spin the width grows linearly in $n$ with a slope that no path optimiser changes (Fig.~\ref{fig:reach}), so the exact methods stop at $n\approx35$--$50$, and on the same families the relaxation of Sec.~\ref{sec:bounds} is not informative beyond $n\approx40$. LABS has width $N-1$ by construction. On these families SA remains the tool, and our results say nothing against it. The $n=150$ 3R3X instance shows the boundary from the other side: reachable on a GPU, but at 7\,h it is not a practical solver, and an attempt at the $L=16$ lattices ($n=256$) was abandoned: the unsliced trees of Table~\ref{tab:reach} cost $2^{43}$--$2^{46}$, but slicing them down to intermediates of $2^{26}$ entries for the GPU inflated the total to $2^{57}$ ($2^{26}\approx67$ million slices of cost $2^{31}$ each), and a 300\,s search found nothing better.

\paragraph{Branch and bound.} A natural next step is to use the slices as subproblems and prune them with the lower bound before contracting. We implemented this and it pruned nothing on the $L=12$ lattices: a slice can be dropped only if its own bound exceeds the incumbent minus one energy unit; the unconditioned bound is 4--8 below the optimum on these instances (Table~\ref{tab:gap}), conditioning on the sliced variables did not close that gap for a single slice (for this test the trees were sliced into 16 to 1024 pieces per instance; every piece had to be contracted), and for instances as degenerate as these almost every slice contains a ground state anyway. Branch-and-bound tensor networks do work elsewhere~\cite{wang2026bbtn,smierzchalski2025spinglasspeps,rams2021approximate}, with branching and bounding schemes that we have not reproduced; our statement is only that slice-level pruning with the generic mini-bucket bound has no value on these instances.

\paragraph{Relation to prior work.} Tropical and generic-semiring contraction are due to~\cite{liu2021tropical,liu2023generic,gao2025guide}, counting with tensor networks to~\cite{kourtis2019fast}, hyper-optimised paths to~\cite{gray2021hyper}, compressed contraction to~\cite{pan2020contracting,gray2024hyperoptimized}, mini-buckets to~\cite{dechter2003minibuckets}. What is new here is modest and specific: the application of all of them to one HOBO object with higher-order terms of any order; the observation that the term-wise network, not the coefficient tensor, is the right input for relaxations; and the use of the exact equilibrium as a reference for SA, which yields the equilibrium-versus-dynamics diagnosis of quadratization and the fairness measurement.

\paragraph{Quadratization as a tensor decomposition.} In the $(+,\times)$ picture a $k$-body Boltzmann factor $e^{-\beta c\prod_{i\in S}x_i}=1+(e^{-\beta c}-1)\prod_{i\in S}x_i$ has CP rank two, and in the tropical picture the standard reduction of a negative monomial ($c<0$), $c\prod x_i=\min_{y\in\{0,1\}} c\,y(\sum_{i\in S}x_i-k+1)$~\cite{freedman2005energy}, is a factorisation through one auxiliary index; an auxiliary bit of Rosenberg's reduction is a bond index of dimension two. We only note the correspondence; whether tropical rank bounds the number of auxiliaries needed by any quadratization is a question we leave open.

\section{Conclusion}
\label{sec:conclusion}

The coefficient tensor of v1 is a definition; the solver is the contraction of the term-wise factor network, and the reduction operator of that contraction is a free parameter. Choosing it as $\min$ solves the problem exactly, as a counting semiring counts the solutions, as a truncated polynomial gives the low-energy spectrum, as $\mathrm{logsumexp}$ gives the free energy, and a top-down pass samples---all from one contraction tree, on lattice instances with $n=144$ variables and up to $4.2\times10^{10}$ ground states within minutes on a CPU, and on one $n=150$ XORSAT instance in 7.3 hours on a GPU. Where exact contraction is out of reach, a width-bounded contraction of the same network gives certified optimality gaps---and on 14 of 44 instances a certificate of optimality---provided the network is built from the problem's own terms; on dense random instances the gaps stay wide. And where neither is possible, the exact quantities on smaller instances serve as a yardstick for the heuristic: on the instances tested they show that quadratization changes the equilibrium weight of the ground states by at most a small factor while SA's success collapses, which points to the single-bit dynamics as the cause, and that SA's coverage of degenerate optima is uneven and, in 10 of 12 instances, worsens after quadratization. We make no claim that tensor networks are faster than annealing; on dense random instances beyond a few dozen variables they are not useful, neither as solvers nor as bound providers. What they offer, where the width allows, is answers of a different kind.

\section*{Code Availability}
\label{sec:code}
The solver (\texttt{hobo.py}: coefficient tensor, quadratization, tropical elimination; \texttt{tn.py}: factor network, semirings, contraction along \texttt{cotengra} trees, exact sampling; \texttt{tn\_torch.py}: min-plus execution on CPU/GPU; \texttt{bounds.py}: dual ascent and mini-bucket; \texttt{sa.py}: batched SA; \texttt{instances.py}: generators), the experiment scripts, the raw results from which every table and figure of this paper is generated, and self-tests against brute-force enumeration are available at \url{https://github.com/minatoyuichiro/tn-hobo-solver} under the MIT license.

\section*{Acknowledgments}
We thank Chia-Wei Hsing for his contributions to the tensor network section of v1 and Shoya Yasuda for his support in providing fundamental knowledge and technical assistance.

\bibliographystyle{unsrt}
\bibliography{references}

\end{document}

%% file: tables/tab_basis.tex
\begin{tabular}{lc ccc}
\toprule
instance & $E_0$ & coefficient tensor (bit monomials) & spin monomials & natural factors \\
& & \multicolumn{3}{c}{lower bound: trivial / $i{=}4$ / $i{=}8$ / $i{=}12$} \\
\midrule
3spin n=18 & $-24$ & $-120.0$ / $-48.3$ / $-31.3$ / $-24.0$ & $-36.0$ / $-34.0$ / $-26.0$ / $-24.0$ & $-36.0$ / $-34.0$ / $-26.0$ / $-24.0$ \\
3SAT n=18 & $+1$ & $-49.0$ / $-7.3$ / $-3.7$ / $-0.7$ & $-18.5$ / $-4.4$ / $-1.7$ / $-0.3$ & $+0.0$ / $+0.0$ / $+0.0$ / $+0.0$ \\
LABS N=12 & $+10$ & $-5458.0$ / $-1567.4$ / $-299.7$ / $+10.0$ & $-370.0$ / $-246.0$ / $-38.0$ / $+10.0$ & $-374.0$ / $-242.0$ / $-38.0$ / $+10.0$ \\
plaq L=4 & $-10$ & $-94.0$ / $-47.8$ / $-19.2$ / $-10.0$ & $-16.0$ / $-16.0$ / $-14.0$ / $-10.0$ & $-16.0$ / $-16.0$ / $-14.0$ / $-10.0$ \\
3R3X n=18 & $-18$ & $-58.0$ / $-20.4$ / $-18.0$ / $-18.0$ & $-18.0$ / $-18.0$ / $-18.0$ / $-18.0$ & $-18.0$ / $-18.0$ / $-18.0$ / $-18.0$ \\
\bottomrule
\end{tabular}

%% file: tables/tab_reach.tex
\begin{tabular}{l rrr rr r rr}
\toprule
family & size & $n$ & terms & \multicolumn{2}{c}{width} & $\log_2$ cost & time [s] & $N_0$ \\
& & & & min-fill & cotengra & (best) & (CPU) & \\
\midrule
planted 3R3X & 20 & 20 & 95 & 7--9 & 6--8 & 10.5--11.6 & 0.00--0.00 & $2$--$4$ \\
 & 40 & 40 & 197 & 13--16 & 12--13 & 16.4--16.8 & 0.00--0.01 & $2$ \\
 & 60 & 60 & 292 & 18--22 & 14--18 & 18.6--22.5 & 0.01--0.16 & $1$--$2$ \\
 & 80 & 80 & 394 & 25--31 & 20--22 & 24.9--27.2 & 0.79--5.85 & $1$--$2$ \\
 & 100 & 100 & 496 & 34--35 & 25--27 & 30.6--32.5 & -- & -- \\
 & 150 & 150 & 748 & 53--58 & 39--41 & 44.8--46.7 & -- & -- \\
 & 200 & 200 & 997 & 72--79 & 52--53 & 57.8--58.4 & -- & -- \\
\addlinespace[2pt]
random 3-spin ($m{=}2n$) & 20 & 20 & 142 & 10--11 & 10--11 & 13.0--14.1 & 0.00--0.01 & $1$--$12$ \\
 & 30 & 30 & 228 & 14--16 & 12--15 & 16.4--18.3 & 0.01--0.01 & $2$--$9$ \\
 & 40 & 40 & 315 & 19--20 & 19--20 & 21.5--22.7 & 0.25--0.26 & $1$--$10$ \\
 & 50 & 50 & 391 & 21--25 & 21--23 & 23.9--26.9 & 1.61--7.92 & $2$--$101$ \\
 & 60 & 60 & 475 & 27--31 & 27--30 & 30.5--33.1 & -- & -- \\
 & 80 & 80 & 665 & 36--38 & 35--38 & 39.6--39.7 & -- & -- \\
 & 100 & 100 & 819 & 41--49 & 45--48 & 44.5--52.1 & -- & -- \\
\addlinespace[2pt]
random 3-SAT ($m{=}4.26n$) & 20 & 20 & 164 & 13--15 & 13--14 & 16.0--17.2 & 0.01--0.01 & $4$--$55$ \\
 & 30 & 30 & 273 & 20--22 & 21 & 23.4--24.5 & 0.32--0.87 & $40$--$274$ \\
 & 40 & 40 & 400 & 28--29 & 27--29 & 30.4--31.4 & -- & -- \\
 & 50 & 50 & 518 & 33--34 & 34--35 & 36.1--36.9 & -- & -- \\
 & 60 & 60 & 639 & 40--43 & 40--44 & 42.8--45.6 & -- & -- \\
 & 80 & 80 & 849 & 52--56 & 53--58 & 54.9--58.8 & -- & -- \\
 & 100 & 100 & 1084 & 68--69 & 71--74 & 71.0--71.5 & -- & -- \\
\addlinespace[2pt]
LABS & 10 & 10 & 221 & 9 & 9 & 11.0 & 0.03 & $40$ \\
 & 15 & 15 & 768 & 14 & 15 & 16.0 & 0.01 & $8$ \\
 & 20 & 20 & 1859 & 19 & 20 & 21.0 & 0.13 & $8$ \\
 & 25 & 25 & 3677 & 24 & 25 & 26.0 & 7.71 & $8$ \\
 & 30 & 30 & 6412 & 29 & 30 & 31.0 & -- & -- \\
 & 35 & 35 & 10252 & 34 & 35 & 36.0 & -- & -- \\
 & 40 & 40 & 15384 & 39 & 40 & 41.0 & -- & -- \\
\addlinespace[2pt]
2D plaquette ($L{\times}L$) & 4 & 16 & 144 & 9 & 8 & 12.2 & 0.00--0.00 & $256$--$1280$ \\
 & 6 & 36 & 304 & 17 & 12 & 17.7--17.7 & 0.01--0.01 & $12288$--$49152$ \\
 & 8 & 64 & 561 & 23 & 16--18 & 23.0--23.1 & 0.16--0.24 & $2.62\times 10^{6}$--$1.18\times 10^{7}$ \\
 & 10 & 100 & 874 & 29 & 22 & 28.0--28.1 & 4.78--7.84 & $1.26\times 10^{7}$--$8.81\times 10^{9}$ \\
 & 12 & 144 & 1232 & 35 & 26 & 33.1--33.2 & -- & -- \\
 & 16 & 256 & 2226 & 47 & 36--38 & 43.6--46.4 & -- & -- \\
\addlinespace[2pt]
2D triangle ($L{\times}L$) & 4 & 16 & 80 & 9 & 8 & 11.6 & 0.00--0.00 & $1$ \\
 & 6 & 36 & 180 & 17 & 12 & 16.7--16.7 & 0.00--0.01 & $16$--$864$ \\
 & 8 & 64 & 320 & 21 & 17--18 & 22.3--22.3 & 0.14--0.15 & $1$ \\
 & 10 & 100 & 500 & 29 & 21 & 27.4--27.5 & 4.86--7.48 & $1$ \\
 & 12 & 144 & 720 & 37 & 25--27 & 32.1--32.8 & -- & -- \\
 & 16 & 256 & 1280 & 45 & 36--38 & 42.6--45.4 & -- & -- \\
\bottomrule
\end{tabular}

%% file: tables/tab_lattice.tex
\begin{tabular}{lrr rrr r rr rr}
\toprule
family & $L$ & $n$ & width & $\log_2$ cost & slices & TN time [s] & $E_0$ & $N_0$ & SA best & SA $P(E_0)$ \\
\midrule
2D plaquette & 12 & 144 & 25 & 32.8 & 2 & 202 & $-136$ & $2.01\times 10^{8}$ & $-136$ & 0.009 \\
2D plaquette & 12 & 144 & 26 & 32.8 & 1 & 219 & $-130$ & $4.23\times 10^{10}$ & $-130$ & 0.032 \\
2D triangle & 12 & 144 & 25 & 32.5 & 2 & 128 & $-140$ & $6912$ & $-140$ & 0.004 \\
2D triangle & 12 & 144 & 26 & 32.4 & 1 & 144 & $-140$ & $13824$ & $-140$ & 0.008 \\
\bottomrule
\end{tabular}

%% file: tables/tab_xorsat.tex
\begin{tabular}{rrl rrr r rr rr rr}
\toprule
$n$ & seed & device & width$^\dagger$ & $\log_2$ cost & slices & TN time [s] & $N_0$ (TN) & $N_0$ (GF(2)) & \multicolumn{2}{c}{SA, $10^3$ sweeps} & \multicolumn{2}{c}{SA, $10^4$ sweeps} \\
& & & & & & & & & best & $P(E_0)$ & best & $P(E_0)$ \\
\midrule
60 & 0 & CPU & 16 & 20.7 & 1 & 0.1 & 1 & 1 & $-58$ & 0.000 & $-60$ & 0.018 \\
60 & 1 & CPU & 14 & 18.6 & 1 & 0.0 & 1 & 1 & $-60$ & 0.002 & $-60$ & 0.022 \\
60 & 2 & CPU & 19 & 22.5 & 1 & 0.1 & 2 & 2 & $-60$ & 0.003 & $-60$ & 0.030 \\
80 & 0 & CPU & 22 & 27.1 & 1 & 4.1 & 1 & 1 & $-80$ & 0.001 & $-80$ & 0.004 \\
80 & 1 & CPU & 23 & 27.0 & 1 & 2.1 & 2 & 2 & $-80$ & 0.001 & $-80$ & 0.005 \\
80 & 2 & CPU & 19 & 24.9 & 1 & 0.5 & 2 & 2 & $-78$ & 0.000 & $-80$ & 0.003 \\
100 & 0 & CPU & 26 & 30.9 & 1 & 49.1 & 2 & 2 & $-98$ & 0.000 & $-100$ & 0.003 \\
100 & 1 & CPU & 24 & 30.2 & 2 & 26.0 & 2 & 2 & $-98$ & 0.000 & $-98$ & 0.000 \\
100 & 2 & CPU & 26 & 32.2 & 2 & 190.1 & 2 & 2 & $-100$ & 0.001 & $-100$ & 0.002 \\
150 & 0 & GPU & 26 & 47.1 & 32768 & 26384.2 & 4 & 4 & $-146$ & 0.000 & $-146$ & 0.000 \\
\bottomrule
\end{tabular}

%% file: tables/tab_quad.tex
\begin{tabular}{lrr r c r c c c c}
\toprule
family & $n$ & seed & aux. & width & $N_0$ & $N(E\le E_0{+}2)$ & $P_0(\beta_{\rm cold})$ & $P_0(2\beta_{\rm cold})$ & SA reaches $E_0$ \\
& & & bits & \multicolumn{6}{c}{each cell: native / $M{=}A{+}1$ / $M{=}1.5A$} \\
\midrule
random 3-spin & 16 & 0 & 18 & 8 / 12 / 13 & 1 & 6 / 6 / 6 & 0.951 / 0.948 / 0.951 & 1.000 / 0.999 / 1.000 & 0.846 / 0.101 / 0.053 \\
random 3-spin & 16 & 1 & 17 & 10 / 15 / 12 & 8 & 61 / 63 / 61 & 0.936 / 0.912 / 0.936 & 0.999 / 0.997 / 0.999 & 1.000 / 0.216 / 0.130 \\
random 3-spin & 16 & 2 & 17 & 9 / 11 / 12 & 2 & 6 / 6 / 6 & 0.978 / 0.978 / 0.978 & 1.000 / 1.000 / 1.000 & 0.846 / 0.045 / 0.033 \\
random 3-spin & 24 & 0 & 33 & 14 / 16 / 17 & 10 & 36 / 63 / 36 & 0.974 / 0.837 / 0.974 & 1.000 / 0.985 / 1.000 & 0.944 / 0.079 / 0.017 \\
random 3-spin & 24 & 1 & 31 & 13 / 19 / 16 & 2 & 20 / 21 / 20 & 0.913 / 0.868 / 0.913 & 0.999 / 0.994 / 0.999 & 0.420 / 0.014 / 0.008 \\
random 3-spin & 24 & 2 & 28 & 12 / 16 / 15 & 1 & 3 / 6 / 3 & 0.980 / 0.809 / 0.979 & 1.000 / 0.980 / 1.000 & 0.673 / 0.023 / 0.002 \\
random 3-spin & 32 & 0 & 44 & 17 / 18 / 17 & 1 & 1 / 2 / 1 & 0.999 / 0.908 / 0.999 & 1.000 / 0.990 / 1.000 & 0.261 / 0.004 / 0.000 \\
random 3-spin & 32 & 1 & 49 & 17 / 19 / 20 & 1 & 16 / 19 / 16 & 0.865 / 0.716 / 0.865 & 0.999 / 0.979 / 0.999 & 0.225 / 0.004 / 0.000 \\
random 3-spin & 32 & 2 & 45 & 17 / 18 / 18 & 8 & 53 / 88 / 53 & 0.944 / 0.756 / 0.944 & 0.999 / 0.977 / 0.999 & 0.671 / 0.006 / 0.002 \\
random 3-spin & 40 & 0 & 59 & 19 / 22 / 22 & 1 & 4 / 7 / 4 & 0.967 / 0.799 / 0.967 & 1.000 / 0.980 / 1.000 & 0.089 / 0.000 / 0.000 \\
random 3-spin & 40 & 1 & 61 & 18 / 22 / 22 & 2 & 29 / 31 / 29 & 0.875 / 0.790 / 0.875 & 0.999 / 0.989 / 0.999 & 0.096 / 0.000 / 0.000 \\
random 3-spin & 40 & 2 & 60 & 19 / 22 / 21 & 10 & 214 / 242 / 214 & 0.824 / 0.683 / 0.824 & 0.998 / 0.981 / 0.998 & 0.303 / 0.001 / 0.001 \\
random 3-SAT & 16 & 0 & 29 & 13 / 16 / 15 & 4 & 185 / 242 / 238 & 0.931 / 0.926 / 0.902 & 0.999 / 0.999 / 0.997 & 1.000 / 0.291 / 0.199 \\
random 3-SAT & 16 & 1 & 29 & 12 / 18 / 19 & 8 & 711 / 1314 / 1947 & 0.861 / 0.827 / 0.650 & 0.998 / 0.998 / 0.972 & 1.000 / 0.292 / 0.213 \\
random 3-SAT & 16 & 2 & 27 & 13 / 18 / 17 & 56 & 1321 / 2666 / 3589 & 0.942 / 0.919 / 0.758 & 0.999 / 0.999 / 0.978 & 1.000 / 0.610 / 0.585 \\
random 3-SAT & 24 & 0 & 43 & 19 / 20 / 19 & 4 & 1550 / 2118 / 2753 & 0.727 / 0.706 / 0.557 & 0.997 / 0.996 / 0.972 & 1.000 / 0.026 / 0.023 \\
random 3-SAT & 24 & 1 & 51 & 18 / 19 / 18 & 18 & 2923 / 4376 / 6094 & 0.871 / 0.833 / 0.626 & 0.999 / 0.998 / 0.967 & 0.969 / 0.022 / 0.015 \\
random 3-SAT & 24 & 2 & 53 & 20 / 19 / 20 & 62 & 4379 / 9939 / 23646 & 0.897 / 0.851 / 0.549 & 0.999 / 0.998 / 0.953 & 1.000 / 0.108 / 0.082 \\
random 3-SAT & 32 & 0 & 70 & 22 / 22 / 22 & 170 & 45977 / 84683 / 167370 & 0.741 / 0.701 / 0.457 & 0.997 / 0.996 / 0.953 & 1.000 / 0.028 / 0.015 \\
random 3-SAT & 32 & 1 & 78 & 22 / 22 / 22 & 10 & 7670 / 11617 / 20314 & 0.651 / 0.609 / 0.392 & 0.995 / 0.995 / 0.954 & 0.979 / 0.003 / 0.003 \\
random 3-SAT & 32 & 2 & 68 & 21 / 22 / 22 & 2 & 8818 / 10413 / 12049 & 0.269 / 0.261 / 0.210 & 0.978 / 0.978 / 0.968 & 0.680 / 0.000 / 0.003 \\
LABS & 10 & 0 & 33 & 10 / 15 / 16 & 40 & 40 / 61 / 40 & 0.322 / 0.195 / 0.321 & 0.673 / 0.466 / 0.673 & 1.000 / 0.097 / 0.086 \\
LABS & 13 & 0 & 63 & 13 / 22 / 22 & 4 & 4 / 13 / 4 & 0.078 / 0.025 / 0.078 & 0.461 / 0.177 / 0.460 & 0.727 / 0.000 / 0.000 \\
\bottomrule
\end{tabular}

%% file: tables/tab_fair.tex
\begin{tabular}{rrr rrrr rrr rr}
\toprule
$n$ & seed & $N_0$ & \multicolumn{4}{c}{native SA} & \multicolumn{3}{c}{quadratized SA} & $H/\log N_0$ native & $p$ \\
& & & hits & max/min & $H/\log N_0$ & $\chi^2$ $p$ & hits & max/min & $H/\log N_0$ & at matched hits & \\
\midrule
24 & 0 & 10 & 15528 & 1.9 & 0.993 & $<10^{-3}$ & 2874 & 7.8 & 0.916 & 0.992 & 0.000 \\
24 & 8 & 9 & 15297 & 1.5 & 0.996 & $<10^{-3}$ & 2562 & 2.9 & 0.974 & 0.995 & 0.000 \\
24 & 10 & 5 & 13519 & 1.2 & 0.998 & $<10^{-3}$ & 969 & 5.8 & 0.884 & 0.997 & 0.000 \\
24 & 17 & 17 & 16070 & 2.4 & 0.986 & $<10^{-3}$ & 2216 & 6.0 & 0.927 & 0.985 & 0.000 \\
32 & 2 & 8 & 10546 & 3.9 & 0.930 & $<10^{-3}$ & 378 & 12.0 & 0.795 & 0.926 & 0.000 \\
32 & 8 & 9 & 10503 & 2.3 & 0.983 & $<10^{-3}$ & 443 & 19.3 & 0.827 & 0.979 & 0.000 \\
32 & 9 & 21 & 13117 & 5.3 & 0.961 & $<10^{-3}$ & 406 & 6.6 & 0.958 & 0.953 & 0.717 \\
32 & 17 & 6 & 8237 & 1.7 & 0.988 & $<10^{-3}$ & 418 & 6.8 & 0.935 & 0.985 & 0.000 \\
40 & 2 & 10 & 4677 & 3.3 & 0.959 & $<10^{-3}$ & 44 & 15.0 & 0.822 & 0.911 & 0.014 \\
40 & 5 & 4 & 2450 & 1.2 & 0.999 & 0.026 & 48 & 1.2 & 0.997 & 0.976 & 0.963 \\
40 & 7 & 11 & 5476 & 3.5 & 0.971 & $<10^{-3}$ & 46 & 9.0 & 0.895 & 0.924 & 0.180 \\
40 & 10 & 8 & 6596 & 3.7 & 0.952 & $<10^{-3}$ & 82 & 7.0 & 0.930 & 0.931 & 0.452 \\
\addlinespace[2pt]
40$^\dagger$ & 2 & 10 & 4706 & 3.4 & 0.958 & $<10^{-3}$ & 653 & 19.9 & 0.864 & 0.955 & 0.000 \\
40$^\dagger$ & 5 & 4 & 2525 & 1.2 & 0.998 & 0.006 & 721 & 1.1 & 1.000 & 0.997 & 0.972 \\
40$^\dagger$ & 7 & 11 & 5472 & 3.5 & 0.966 & $<10^{-3}$ & 817 & 18.0 & 0.871 & 0.964 & 0.000 \\
40$^\dagger$ & 10 & 8 & 6548 & 3.5 & 0.948 & $<10^{-3}$ & 1248 & 8.2 & 0.904 & 0.946 & 0.000 \\
\bottomrule
\end{tabular}

%% file: tables/tab_gap.tex
\begin{tabular}{lrrr r rrr rr}
\toprule
family & size & seed & $n$ & UB (SA) & \multicolumn{3}{c}{natural factors} & \multicolumn{2}{c}{coefficient tensor} \\
& & & & & dual & LB & gap & LB & gap \\
\midrule
planted 3R3X & 40 & 0 & 40 & $-40$ & $-40.0$ & $-40.0$ & 0$^*$ & $-40.0$ & 0$^*$ \\
planted 3R3X & 40 & 1 & 40 & $-40$ & $-40.0$ & $-40.0$ & 0$^*$ & $-40.0$ & 0$^*$ \\
planted 3R3X & 60 & 0 & 60 & $-60$ & $-60.0$ & $-60.0$ & 0$^*$ & $-60.0$ & 0$^*$ \\
planted 3R3X & 60 & 1 & 60 & $-60$ & $-60.0$ & $-60.0$ & 0$^*$ & $-60.0$ & 0$^*$ \\
planted 3R3X & 100 & 0 & 100 & $-96$ & $-100.0$ & $-100.0$ & 4 & $-100.0$ & 4 \\
planted 3R3X & 100 & 1 & 100 & $-98$ & $-100.0$ & $-100.0$ & 2 & $-101.8$ & 2 \\
planted 3R3X & 200 & 0 & 200 & $-192$ & $-200.0$ & $-200.0$ & 8 & $-200.8$ & 8 \\
planted 3R3X & 200 & 1 & 200 & $-194$ & $-200.0$ & $-200.0$ & 6 & $-203.4$ & 8 \\
random 3-spin & 40 & 0 & 40 & $-60$ & $-80.0$ & $-60.0$ & 0$^*$ & $-60.0$ & 0$^*$ \\
random 3-spin & 40 & 1 & 40 & $-58$ & $-80.0$ & $-58.0$ & 0$^*$ & $-58.0$ & 0$^*$ \\
random 3-spin & 60 & 0 & 60 & $-88$ & $-120.0$ & $-104.0$ & 16 & $-109.2$ & 20 \\
random 3-spin & 60 & 1 & 60 & $-88$ & $-120.0$ & $-102.0$ & 14 & $-113.6$ & 24 \\
random 3-spin & 100 & 0 & 100 & $-150$ & $-200.0$ & $-192.0$ & 42 & $-211.4$ & 60 \\
random 3-spin & 100 & 1 & 100 & $-150$ & $-200.0$ & $-195.3$ & 44 & $-217.4$ & 66 \\
random 3-spin & 150 & 0 & 150 & $-226$ & $-300.0$ & $-300.0$ & 74 & $-344.8$ & 118 \\
random 3-spin & 150 & 1 & 150 & $-226$ & $-300.0$ & $-300.0$ & 74 & $-346.1$ & 120 \\
random 3-SAT & 40 & 0 & 40 & $+0$ & $+0.0$ & $+0.0$ & 0$^*$ & $-4.5$ & 4 \\
random 3-SAT & 40 & 1 & 40 & $+1$ & $+0.0$ & $+0.0$ & 1 & $-4.6$ & 5 \\
random 3-SAT & 60 & 0 & 60 & $+1$ & $+0.0$ & $+0.0$ & 1 & $-11.3$ & 12 \\
random 3-SAT & 60 & 1 & 60 & $+1$ & $+0.0$ & $+0.0$ & 1 & $-13.1$ & 14 \\
random 3-SAT & 100 & 0 & 100 & $+0$ & $+0.0$ & $+0.0$ & 0$^*$ & $-26.2$ & 26 \\
random 3-SAT & 100 & 1 & 100 & $+0$ & $+0.0$ & $+0.0$ & 0$^*$ & $-26.5$ & 26 \\
random 3-SAT & 150 & 0 & 150 & $+0$ & $+0.0$ & $+0.0$ & 0$^*$ & $-34.9$ & 34 \\
random 3-SAT & 150 & 1 & 150 & $+1$ & $+0.0$ & $+0.0$ & 1 & $-30.3$ & 31 \\
2D plaquette & 8 & 0 & 64 & $-54$ & $-64.0$ & $-54.0$ & 0$^*$ & $-68.4$ & 14 \\
2D plaquette & 8 & 1 & 64 & $-56$ & $-64.0$ & $-56.0$ & 0$^*$ & $-70.7$ & 14 \\
2D plaquette & 12 & 0 & 144 & $-136$ & $-144.0$ & $-140.0$ & 4 & $-211.5$ & 74 \\
2D plaquette & 12 & 1 & 144 & $-130$ & $-144.0$ & $-138.0$ & 8 & $-207.0$ & 76 \\
2D plaquette & 16 & 0 & 256 & $-234$ & $-256.0$ & $-246.0$ & 12 & $-341.1$ & 106 \\
2D plaquette & 16 & 1 & 256 & $-236$ & $-256.0$ & $-248.0$ & 12 & $-345.6$ & 108 \\
2D plaquette & 20 & 0 & 400 & $-358$ & $-400.0$ & $-400.0$ & 42 & -- & -- \\
2D plaquette & 20 & 1 & 400 & $-358$ & $-400.0$ & $-400.0$ & 42 & -- & -- \\
2D plaquette & 24 & 0 & 576 & $-508$ & $-576.0$ & $-568.0$ & 60 & -- & -- \\
2D plaquette & 24 & 1 & 576 & $-510$ & $-576.0$ & $-570.0$ & 60 & -- & -- \\
2D triangle & 8 & 0 & 64 & $-64$ & $-64.0$ & $-64.0$ & 0$^*$ & -- & -- \\
2D triangle & 8 & 1 & 64 & $-64$ & $-64.0$ & $-64.0$ & 0$^*$ & -- & -- \\
2D triangle & 12 & 0 & 144 & $-140$ & $-144.0$ & $-144.0$ & 4 & -- & -- \\
2D triangle & 12 & 1 & 144 & $-140$ & $-144.0$ & $-144.0$ & 4 & -- & -- \\
2D triangle & 16 & 0 & 256 & $-248$ & $-256.0$ & $-256.0$ & 8 & -- & -- \\
2D triangle & 16 & 1 & 256 & $-246$ & $-256.0$ & $-256.0$ & 10 & -- & -- \\
2D triangle & 20 & 0 & 400 & $-382$ & $-400.0$ & $-400.0$ & 18 & -- & -- \\
2D triangle & 20 & 1 & 400 & $-382$ & $-400.0$ & $-400.0$ & 18 & -- & -- \\
2D triangle & 24 & 0 & 576 & $-546$ & $-576.0$ & $-576.0$ & 30 & -- & -- \\
2D triangle & 24 & 1 & 576 & $-548$ & $-576.0$ & $-576.0$ & 28 & -- & -- \\
\bottomrule
\end{tabular}